\documentclass[11pt]{article}
\usepackage{geometry}                
\usepackage{graphicx}
\usepackage{amssymb}
\usepackage{amsmath}
\usepackage{fixmath}
\usepackage{MnSymbol}
\usepackage{amsfonts}
\usepackage{array}
\usepackage{multirow}
\DeclareMathAlphabet{\mathpzc}{OT1}{pzc}{m}{it}

\usepackage{caption}
\usepackage{subcaption}

\title{Surface Shear and  Persistent Wave Groups}
\author{Clifford Chafin\\\ \small{Department of Physics, North Carolina State University, Raleigh, NC 27695} \thanks{cechafin@ncsu.edu}}

\begin{document}
\maketitle

\begin{abstract}
We investigate the interaction of waves with surface flows by considering the full set of conserved quantities, subtle but important surface elevation changes induced by wave packets and by directly considering the necessary forces to prevent packet spreading in the deep water limit.  Narrow surface shear flows are shown to exert strong localizing and stabilizing forces on wavepackets to maintain their strength and amplify their intensity even in the linear regime.  Subtle packet scale nonlinear elevation changes from wave motion are crucial here and it suggest that popular notions of wave stress and action are naive.  Quantitative bounds on the surface shear flow necessary to stabilize packets of any wave amplitude are given.  One implication of this mechanism is that rogue wave stabilization must be due to a purely nonperturbative process.  
\end{abstract}

Rogue waves or ``monster waves'' are anomalously large waves that appear at sea and in very large lakes.  Dismissed as sailors lore until the mid 1990's these waves appear in isolation or groups of three and may be a few times larger than the surrounding waves sometimes reaching over 30m.  The damage and destruction these waves have delivered to the largest vessels and oil platforms is astounding.  Aside from their scale, the striking feature of these waves is their frequency.  They occur far more often than statistics would predict and their great difference in scale relative to the surrounding waves suggests some new mechanism is at work \cite{Review}.  Recently there has been success in generating optical analogs to rogue waves but the understanding of the oceanic version has been elusive.

The interaction of flows and waves is of interest for a number of reasons.  It has been suggested that surface shear triggers rogue waves \cite{Shear}.  The appearance of whale ``footprints'' \cite{Levy} from circulatory fluke motion that leads to persistent calming of water indicates that shear can have a powerful effect on waves.  It has been observed that real ocean waves are never purely irrotational but contain ``Eulerian'' shear flows that tend to exactly cancel the Stokes drift predicted for waves.  Experimental measurements show this is a universal effect for wavemakers but the details can vary wildly depending on the implementation.  Results in ocean measurements \cite{Smith} show that persistent wave groups also appear.  It is unknown if the surface shear contributes to this effect.  

The study of surface gravity waves on an incompressible fluid is a very old subject of classical mechanics \cite{Stokes, Gerstner}.  Despite this, there are many unresolved problems and a feeling by many members of the field that it is time for a reconsideration of many or all topics of established theory \cite{Ca07}.  The theory of simple fluids has been established since the 19th century.  It is a testament to the difficulties of nonlinear and continuum mechanics that any significant mysteries about surface waves remain.  Some interesting results have been obtained by the use of ``wave stress'' and ``wave action.''  Through a careful reconsideration of the totality of conserved quantities for waves we will see that the first leads to some paradoxes which, with some more consideration, are fixable.  The later concept, wave action, turns out to have destructive inconsistencies that casts serious doubt on its usefulness, and that of the associated generalized Lagrangian mean approach, doubtful.  The usual momentum/pseudomomentum confusion raises its ugly head here along with some of the subtle ways boundary terms can make surprising contributions to conservation laws for harmonic solutions.  

There is some extensive literature on rogue waves and their generation by way of perturbative treatments of hydrodynamics usually leading to the nonlinear Schr\"{o}dinger equation \cite{Ak}.  This treatment is meant to be complimentary to the treatments that rely on shear flow and to cast doubt on those that do not.  Our treatment will be based on the mechanical forces induced on packets necessary to produce packets without dispersion and to concentrate energy in them.  The manner in which nonlinear behavior can be included perturbatively can be done in a rather formulaic manner but sometimes these effects are eliminated either by higher order considerations or because the presumed basis set is inadequate to give the results by any perturbative means.  An advantage of conserved quantities is they are always rigidly conserved but there can be subtleties to the locality of this process in continuum mechanics.  For this reason, the bulk of this paper is dedicated to a careful discussion of these conservation laws.  The final results then appear clearly and convincingly from them.  

The outline of this paper is as follows.  We will begin with a brief review of Airy waves, the conservation laws then introduce some of the subtleties of boundary effects and the necessity of wave packet analysis.  The notion of momentum flux, waves stress and Reynold's stresses are shown to give a paradox unless nonlinear pressure driven surface elevation is correctly accounted for.  This effect is essential no matter how far into the ``linear regime'' we make the system.  Finally, we use these effects to compute the action of a thin surface shear on a nearly monochromatic wavepacket and show that it can lead to both stabilization of the packet against dispersive spreading and can cause compression of it to create higher energy densities depending on whether the flow and the waves are co- or counter-moving.  Such an effect exists in the limit of linear waves where no higher Stokes corrections are necessary.  The delineation of what ``linear'' should mean for surface waves is complicated but, since we will account for all conserved quantities: energy, momentum, angular momentum, mass and vorticity, this discussion should be considered a more authoritative approach to the problem than previous arguments involving the nonlinear Schr\"{o}dinger equation and other perturbative methods.  

\section{Basis of Airy Waves}\label{Small}

The Navier-Stokes equations govern wave evolution.  This can generally be chosen to be for the case of a fluid where the density of each parcel is constant and the pressure is propagated instantaneously.  These approximations are excellent since, although we consider the deep water limit, the wavelength in practice is small enough that the wave speed is always much less than the sound velocity and diffusion and migration between depths is very slow.  The resulting equations are
\begin{align*}
\frac{\partial}{\partial t}v+v\cdot\nabla v=-\frac{1}{\rho}P+gz+\nu\nabla^{2}v
\end{align*}

The pressure inside a wave is generally not a harmonic function.  It satisfies $\nabla^{2}P=-\rho\partial_{i}v_{j}\partial_{j}v_{i}$.  Consider the case of a a water surface with some elevated regions but instantaneously no velocity and a wind blowing across it exerting surface forces.  In this case, the pressure obeys $\nabla^{2}P=0$.  We can always decompose the internal pressure of a wave into that driven by the inertia of the flow and elevation changes and that given by surface forces $P=P_{i}+P_{s}$.  We see that the (relatively small) surface force contribution to the pressure is always harmonic.  From the N-S equations we have 
\begin{align*}
\frac{\partial}{\partial t}v+v\cdot\nabla v=-\frac{1}{\rho}(\nabla P_{i}+\nabla P_{s})+gz
\end{align*}
This shows that the change in $\rho \partial_{t}v$  over what the wave motion is without the surface forces is $-\nabla P_{s}$.  Since this is a harmonic change, and the whole history of the wave is built up from such changes, we are led to conclude that $v$ itself is always irrotational and all the rotational components of the motion decouple from these forces.  The irrotationality of $v$ means we can introduce a velocity potential $v=\nabla\Phi$.  The irrotationality of waves is generally argued to be true because vorticity is only advected and not transported with the wave.  

For small waves we neglect these higher order terms and derive the Airy theory solutions.  This does give us a solution for every wavelength of a uniformly progressing surface state.  
The fully nonlinear N-S equations yield the Bernoulli equation
\begin{align}
\partial_{t}\Phi+\frac{1}{2}\nabla\Phi\cdot\nabla\Phi -g z+\frac{P_{eq}}{\rho}+\frac{P_{w}}{\rho}=0\\ \nonumber
\end{align}
where $P_{w}$ is the change\footnote{$P_{w}$ is not assumed to be ``small.''} in the pressure from that equilibrium case of a flat surface of water at rest.  
Incompressibility follows from the Poisson equation $\nabla^{2}\Phi=0$.  This acts as a constraint that fixes the pressure.

The definition of ``small'' for periodic waves is simple.  We call a wave with amplitude and wavelength $(a,\lambda)$ small if $a/\lambda<<1$.  In the N-S equations we can see that the nonlinear term $\nabla\Phi\cdot\nabla\Phi$ is smaller than $\partial_{t}\Phi$ and $P$ everywhere in the solution.  As a secondary point, we cannot assume that the linear superposition of two such ``small'' waves keeps this term small.  If we superimpose two small periodic waves $(A,\Lambda)$ and $(a,\lambda)$ where $A>a$ and $\Lambda>\lambda$ but $A\sim\lambda$ we get a cross term that is does not let us decompose the terms into two independent equations.  

During wave creation we assume the net vertical force over the whole surface is zero and there is some bound on the size of region where the forces can be net up or net down.  This is consistent with our understanding of wind driving a wave by exerting pressure on the windward side of the wave and a deficit in the lee.  After creation we assume the pressure is constant and seek a basis for the ``small'' waves as follows.  

Since the pressure and velocity potential in this limit must be harmonic we know they should be oscillatory in one direction and exponential in the other.  This suggests that we use 
the following basis to span the space of all allowed velocity potentials $\Phi(x,z)$ and pressures $P_{w}(x,z)$.
\begin{align}
\phi_{k}(x,z)&=e^{kz}\cos(kx)\\
\phi_{k}^{\dagger}(x,z)&=e^{kz}\sin(kx)\\ \nonumber
\end{align}

In the general case of an arbitrary small surface deformation $\eta(x)$ we want to be able to provide and arbitrary velocity potential $\Phi(x,z)$ and evolve it.  Without a well defined wavelength we need a new criterion for smallness.  We want the nonlinear term to be ignorable so we can choose $\nabla\Phi\cdot\nabla\Phi<<P_{w}$ to hold for all $(x,z)$ beneath the surface of the wave.  In this limit, we can show that the pressure is harmonic and so can be described on the $\Phi$-basis $\{\phi_{k}, \phi_{k}^{\dagger}\}$.  
It is convenient in Airy theory to be able to approximate the values of $\Phi$ and $\nabla\Phi$ at the surface $\eta(x)$ by the values at the equilibrium height $z=0$.

Now let us build up a basis for the general small standing surface waves $(\eta(x),\Phi(x,z))$.  These are the Airy waves \cite{Ai45}.  The general ``small'' waves are only a proper subset of the span of this basis.   This is because we know that Fourier expansions with ``small'' coefficients can give locally large values in the represented function (e.g.\ the $\delta$-function).   We expect sufficiency for this bases by our argument on the allowed forms form of the functions $P_{w}$ and $\Phi$.  Clearly this basis can represent any (single valued i.e.\ not breaking) surface function $\eta(x)$ since it contains all Fourier components.  

We must satisfy the following (linearized) dynamic, kinematic and incompressibility conditions:
\begin{align}\label{eom}
\frac{\partial\Phi}{\partial t}&=-g\eta ~~\text{at z=$\eta(x)$}\\
\frac{\partial \eta}{\partial t}&=\frac{\partial \Phi}{\partial z}~~\text{at z=$\eta(x)$}\nonumber\\
\frac{\partial^{2} \Phi}{\partial x^{2}}&=-\frac{\partial^{2} \Phi}{\partial z^{2}}\nonumber
\end{align}
From the above smallness criterion we can evaluate these, to lowest order, at $z=0$ instead of at $z=\eta$.

We try the following (standing wave) trial solution:
\begin{align}
\Phi(x,z,t)=\phi^{0}\cos(\omega t)=\bigg(\sum_{k}a_{k}\phi_{k}(x,z)\bigg)~\cos(\omega t)
\end{align}
Combining the first two equations we find:
\begin{align*}
\frac{\partial^{2}\Phi}{\partial t^{2}}&=-g\frac{\partial \Phi}{\partial z} ~~\text{at z=0}\\
-\omega^{2}\phi_{k}&=-gk\phi_{k}\\
\end{align*}
with $a_{k}=0$ unless
 the usual dispersion relation holds
\begin{align*}
\omega_{k}=\pm\sqrt{gk}
\end{align*}
From the first equation we find
\begin{align*}
\eta_{k}(x,t)=\frac{k}{\omega}\phi_{k}(x,0)\sin(\omega t)
\end{align*}
This means we can evolve from any initial data of $(\Phi, \eta)$ of the form $(\phi_{k}(x,z), 0)$ or $(0,\frac{k}{\omega}\phi_{k}(x,0))$.  This set (and the corresponding functions for the $\phi_{k}^{\dagger}$ set) spans the set of all possible $\eta(x)$ and $\phi(x,0)$ which means we can evolve any such $(\Phi, \eta)$ we choose with our linearized eom subject to the smallness conditions. 

\section{Current-Flow Decomposition}
A wave that is initiated by a localized impulse, like a falling stone, gives a local turbulent shear and a surface deformation.  In practice the surface deformation generates a pressure field far beyond the local turbulence and it propagates away at a much faster rate than the turbulence diffuses.  This is why waves are generally modeled as irrotational.  The general velocity field can be represented as $v=\nabla \phi+\nabla\times u$, for suitable boundary conditions \cite{Morse}, according to the Helmholtz decomposition theorem.  This is the fashion is which we define a separation of velocity into wave motion and ``currents.''  

A measure of the rotational properties of a flow can therefore be given by the vorticity: $\omega=\nabla\times v$.  The vorticity transport theorem can be expressed
\begin{align*}
\frac{\partial}{\partial t}\omega+v\cdot\nabla \omega= (\omega\cdot\nabla) v -\nu\nabla^{2}\omega
\end{align*}
where $\nu=\eta/\rho$ is the kinematic viscosity.  When the viscous losses are small as they tend to be for waves and gentle shear gradients and there is no ``stretching'' of the flow along the vorticity it says that the vorticity is conserved and advected with the flow.

\section{Conserved Quantities of Plane Waves}\label{Conserved}
The conserved quantities of classical mechanics for a system isolated from its surroundings are mass, energy, momentum, and angular momentum.  If the fluid has no viscosity then vorticity is conserved as well \cite{Batchelor}.  Incompressibility and irrotationality imply that $\nabla^{2}\Phi=0$.  There are other conserved quantities such as the ``wave action'' but we are interested in how waves are created by forces outside of the fluid.  This means we want the universally conserved quantities of classical physics.  These are the ones we know how to relate from one system to another.  Internally conserved quantities such a ``quasi-momentum'' would need some source term (of non-obvious form) when external forces act on the system.  
In the following, we will keep only the lowest order contributions in each case.

The energy density (depth integrated energy per area) comes from kinetic and potential parts\footnote{We choose a right handed coordinate system with +x pointing right, +z pointing up and +y pointing into the page.}:
\begin{align}
\mathcal{E}&=\mathcal{K}+\mathcal{U}=\frac{1}{\lambda}\int_{0}^{\lambda}\int_{-\infty}^{\eta(x)}\frac{1}{2}\rho(v_{x}^{2}+v_{z}^{2})dz~dx
+\frac{1}{\lambda}\int_{0}^{\lambda}\int_{0}^{\eta(\lambda)}(\rho g h)~ dh~ dx\\\nonumber
&=\frac{1}{2}\rho g a^{2}\\\nonumber
\end{align}

The linear momentum density of an Airy has a nonzero component only in the +x direction so can be directly calculated.  
\begin{align}
\mathpzc{p}&=\frac{1}{\lambda}\int_{0}^{\lambda}\int_{-\infty}^{\eta(x)}\rho v_{x} dz~dx\\\nonumber 
&=\frac{1}{2}\rho\omega a^{2}\\\nonumber
\end{align}
There has been concern that there are additional contributions to momentum that are ``hidden'' in that they are fluxes not associated with the motion of the momentum density of the waves.  We will address this thoroughly in following sections.  

Angular momentum density is trickier.  A first naive check is to realize that each particle moves in circle of radius $a e^{kz}$ with velocity $\omega a e^{kz}$.  Integrating gives an angular momentum density of $+\frac{1}{2}\rho g a^{2}/\omega$ \cite{LH80}.  This seems like an exact result for the Gerstner waves based on the parallel axis theorem or a Lagrangian average of the motion of particles in wavelength over a period.\footnote{Gerstner gave an early nonirrotational wave solution where particles exhibit perfect circles with no drift.  These are the only closed form exact solutions of waves known but they only describe waves in a infinite particular shear flow so are not of much physical interest.}   Airy waves would then have a correction due to the Stokes drift.  The potential problem with this is that angular momentum has to be calculated about a fixed point and be done at a \textit{fixed time}.   The most natural base points to use would be ones on the surface of equilibrium height.  So what could go wrong?  
\begin{enumerate}
\item Different base points at the equilibrium surface may give different angular momenta densities for a given limit finite length cutoffs going to infinity.  
\item Small nonlinear pressure driven elevation changes of the packet from the equilibrium surface due to wave motion might create large contributions over the long length of the packet.  
\item  The angular momentum from each wavelength increases linearly with the packet length so these could be of the same order.  This would make the result sensitive to how the packet is truncated or attenuated at the ends.  
\end{enumerate}

The first and last points are essentially the same and will be resolved below.  It will turn out that the surface elevation can make a measurable change in a long packet.  This is something that must be considered for any use where angular momentum is important.  The first and last points require we know how to evaluate the angular momentum density for an infinite wave.  Contributions at infinity can only be resolved by the use of packets.\footnote{This is not a new problem.  There are known examples in electromagnetism where taking the limit to a ``nice'' highly symmetrical example, usually to simplify calculations, introduces some singular complications.  A famous examples is the linear momentum of the EM field for static crossed E and B fields where the linear momentum to cancel the field is hidden at infinity as the ``hidden momentum.''}  

The last point suggests that to get the angular momentum density of an infinite wave we should start with packets and take the limit.\footnote{This is ironic because we normally find quantities like energy density for infinite waves, mathematically ``nice,'' solutions, then apply them to the case of localize packets.  Here we must look at the localized packets to resolve the angular momentum for the plane waves.}  
Let us first consider a set of $N$ wavelengths of a traveling wave and place them in a box with rigid walls as initial data.  If the box is set on a pivot so that it can only rotate about the base point and we turn off gravity and let it evolve until equilibrium is obtained (assuming some small viscosity) we can extract the angular momentum of the wave by the final motion of the container.  This shows that the quantity is experimentally well defined.  

We can directly calculate this for $N$ wavelengths of a right moving wave about $(0, 0)$ at $t=0$:
\begin{align}\label{L}
\mathcal{L}&=\frac{1}{N\lambda}\int_{\pi/2k+\delta/2k}^{\pi/2k+N\lambda+\delta/2k}\int_{-\infty}^{\eta(x)}\rho (z v_{x}-x v_{z}) dz~dx\\
&=-\frac{\omega}{k}\frac{a}{k}\sin\left(\frac{\delta}{2}\right)-\frac{1}{4}\frac{\omega}{k} a^{2}\left( 2\cos^{2}\left(\frac{\delta}{2}\right)-1\right)\nonumber
\end{align}
where we have started the integration at the first zero of $\eta(x)$ and $\delta$ is a phase to shift the interval where we truncate the wave.  It is easy to see, since the $-x v_{z}$ term usually dominates, that shifting the base point on the surface, about where we calculate angular momentum, makes no difference but we see that where we cut the wave matters a lot.  The minimal solution is the one where $\delta=0$ the ends of the packet have $\eta=0$ which is where the $z v_{x}$ term dominates.  For an actual packet this is the most reasonable choice.  The packet attenuates at the ends to the background sea level not an abrupt step in elevation.  Of course, we can choose any initial data we choose but this choice is the one a packet will tend to relax to with higher order components separating spatially from it or resulting in wave breaking that transfers momentum to slower shear flow.  As we change $\delta$ we pickup \textit{lower order} contributions but the net average of these makes the angular momentum negative.  This is \textit{opposite} the direction of particle rotation, our above estimate and the results of standard theory\footnote{Longuet-Higgins uses an x-y instead of an x-z basis, as we do here, but also uses a nonstandard definition of angular momentum in that motions that makes cw motion give positive values.  He takes pains to calculate $\mathcal{L}$ from Lagrangian and Eulerian points of view.  Our result agrees with his Eulerian calculation.  Only the Eulerian angular momentum is the physical one that is conserved under interactions with the external world.}\cite{LH80} based on the time averages of Lagrangian motion.  This is, of course, disturbing.  We should be able to probe angular momentum of the wave during wave destruction at the end of a tank.  The Stokes drift of the wave must crash into the wall and transfer momentum to it as a counterrotation but this is supposed to be a smaller contribution to the angular momentum than the forwards particle rotation.  It immediately makes us concerned that the Lagrangian average of particle motion in an infinite wave may not be the true conserved angular momentum that we evaluate at any given \textit{instant} in time.  
Our direct computation 
specifies the angular momentum of a packet as 
\begin{align}
\mathcal{L}=-\frac{1}{4}\rho\frac{\omega}{k} a^{2}=-\frac{1}{4}\rho\frac{g a^{2}}{\omega}  
\end{align}
When we investigate angular momentum of packets and their means of creation it will be clear that this is the necessary sign of angular momentum for a surface wave.  Specifically, right moving waves give counterclockwise angular momentum.

Computing momentum flux is more subtle than that of mass flux (momentum density).  Mass is a locally conserved quantity so is only advected.  Momentum is created and destroyed at pressure gradients.  
Angular momentum is even more subtle and different from the other conserved quantities in requiring a basepoint to compute it about.  The fact that this quantity scales linearly with the distance from this point will be why small changes at the ends of wavepackets can dominate and swamp the entire contribution of the finite wavetrain.  

If we instead calculate the angular momentum about a point deep under the waves we see that the sign is positive as we originally expected.  Interestingly the dominant terms of $L$ come from the $z v_{x}$ term instead of the $-x v_{z}$ term as above.  This makes the role of the Stokes drift the dominant contribution.  Of course, we only care about angular momentum in relation to torques which must be calculated about the new base point as well.  
For a packet that is narrow compared to the depth of the base point, the vertical motions are now higher order and the Stokes drift dominates.  This means that $L\sim R\times p_{\text{stokes}}$.  In the limit of large $R$ this reduces the number of independent equations we infer from the lowest order forces and torques on the waves from two to only one simply because of the change in base point.\footnote{Angular momentum and the center of mass moment are rather special among the classical conservation laws since they depend on the the relative location to a specified base point, not just local intrinsic properties.  When evaluating $L$ for this packet as $R\rightarrow\infty$ we have to be careful about how rapidly the length of the wavetrain $l$ grows in relation to $R$.  By assuming $l<<R$ as $l\rightarrow\infty$ we get the result quoted here and that the $v_{x}$ motion is the only  one that contributes.  Such an ambiguity is not comforting  and reinforces our notion that the surface is the most natural choice for this geometry. }   Since we are interested in applying conservation laws to constrain evolution we want as many as possible.  Therefore, for the flat sea case, the better choice is to put the base point on the surface.

The conserved quantities for the Airy waves are summarized in table \ref{tab}.  The relevance of these for packet motion will be discussed next.\footnote{The mass density in the following table will be derived and explained in Sec.~\ref{Packets} after we do some further investigation of packet motion.}  
\begin{table}
\begin{center}
\vspace{0.5cm}
\begin{tabular}{|l|l|l|l|}  \hline $\mathcal{E} $& $\mathit{m}$&$\mathpzc{p}$ & $\mathcal{L}$ \\  \hline \multirow{2}{*}{$\frac{1}{2}\rho g a^{2}$}& \multirow{2}{*}{$\rho a^{2}  k$}  & \multirow{2}{*}{$\frac{1}{2}\rho a^{2} \omega $} &\multirow{2}{*}{$-\frac{1}{4}\rho g \frac{ a^{2}}{\omega}$}  \\ &  &  &\\ \hline \end{tabular} 
\vspace{0.5cm}
\end{center}
\caption{Conserved depth-integrated and time averaged quantities.}\label{tab}
\end{table}

\section{Conservation Laws and Packets}\label{Packets}
\begin{figure}[!h]
   \centering
   \includegraphics[width=4in,trim=0mm 160mm 0mm 120mm,clip]{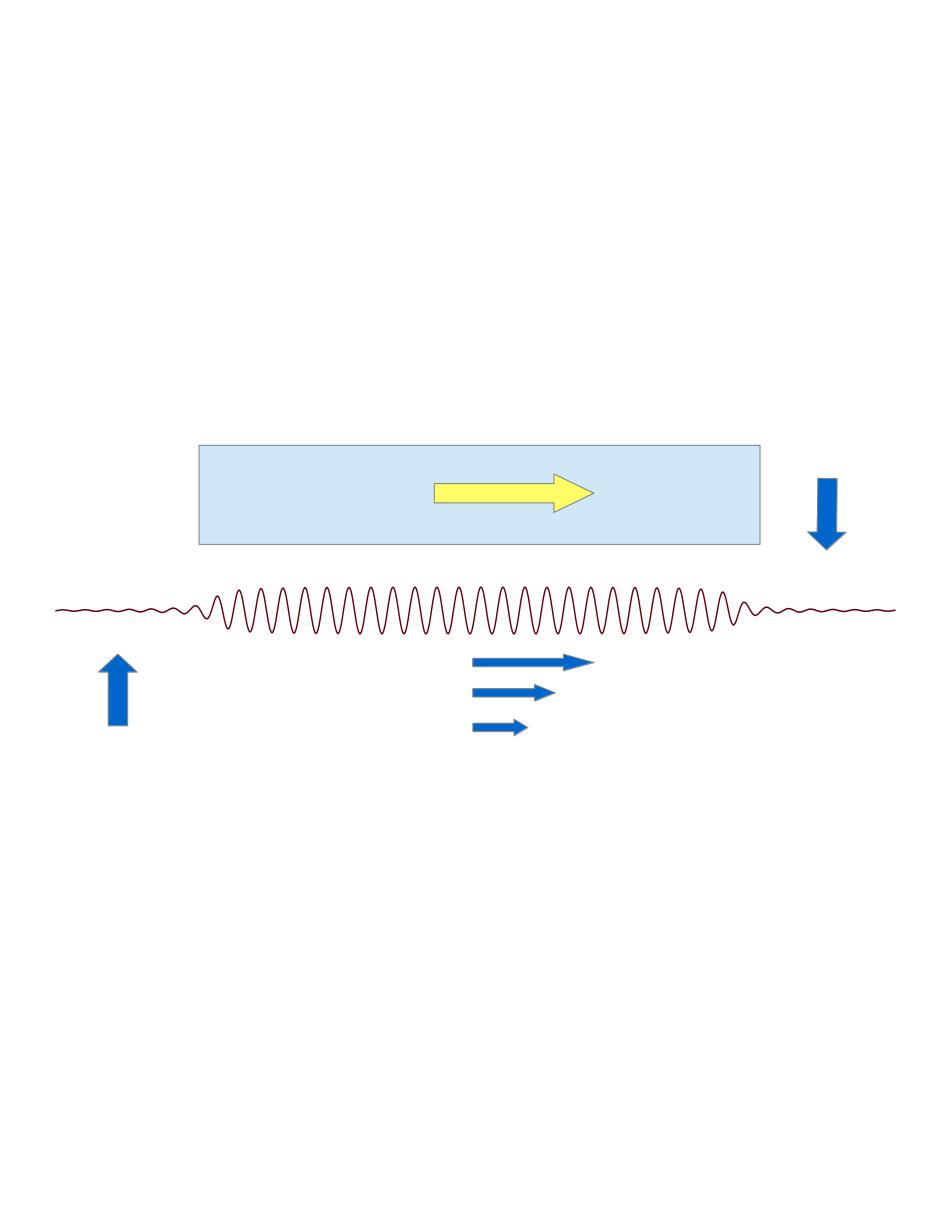} 
   \caption{A wind parcel trades linear and angular momentum with the sea surface producing waves, surface shear and end of packet elevation changes.  The surface waves are assumed small in the sense that $ak\ll 1$ though the elevation is exaggerated for illustration.}
   \label{windshear}
\end{figure}

Let us now consider the way surface waves are generally generated on the sea, by wind.  Consider a large parcel of air as in fig.\ \ref{windshear} as might be driven by a pressure gradient over finite segment of ocean.  The original angular momentum measured from point on the surface is clockwise (cw) i.e.\ positive since only the air is moving.  Analogously to the case of a car braking on pavement, the drag under the parcel is forwards so transfers a counterclockwise (ccw) momentum to the water.  This is made up for by the forwards pitching of the car, or air parcel in our case, that preservers the net cw angular momentum and gives a leading depression and trailing rise on our wavetrain that proceed to rapidly leave the slower wavetrain behind due to their much faster group velocity.  

At this point, we note that the ``wave action'' $\mathcal{E}/\omega$ is proportional to the angular momentum density.  The bulk of angular momentum is in long waves that are on the order of the length of the wavetrain and independent of the local wavelength.  The ends of the packet tend to spread through dispersion, and small elevation changes at transitions can create a long lasting torque.  Furthermore, the angular momentum is not a locally advected quantity and contributions far away from the basepoint are disproportionately weighted.  It is independent of energy and linear momentum which all must be simultaneously conserved.  This makes the value of wave action as a locally valuable quantity that can determine wave interactions also questionable.  When we consider the attenuation of the waves from damping we see that there is only a remaining forwards shear so the angular momentum of it must be ccw confirming the sign in our earlier calculation.  Considering the attenuation of an infinite monochromatic\footnote{i.e.\ single frequency.} surface wave we know that the mass loss is into a surface shear since this is where the vorticity must enter the fluid but where does the angular momentum go?  In the case of a packet we see that it is ejected into irrotational surface elevation changes at the ends of the packet.  This gives more reason to be very skeptical of local use of the wave action.  

\begin{figure}[!h]
   \centering
   \includegraphics[width=3in,trim=0mm 165mm 0mm 90mm,clip]{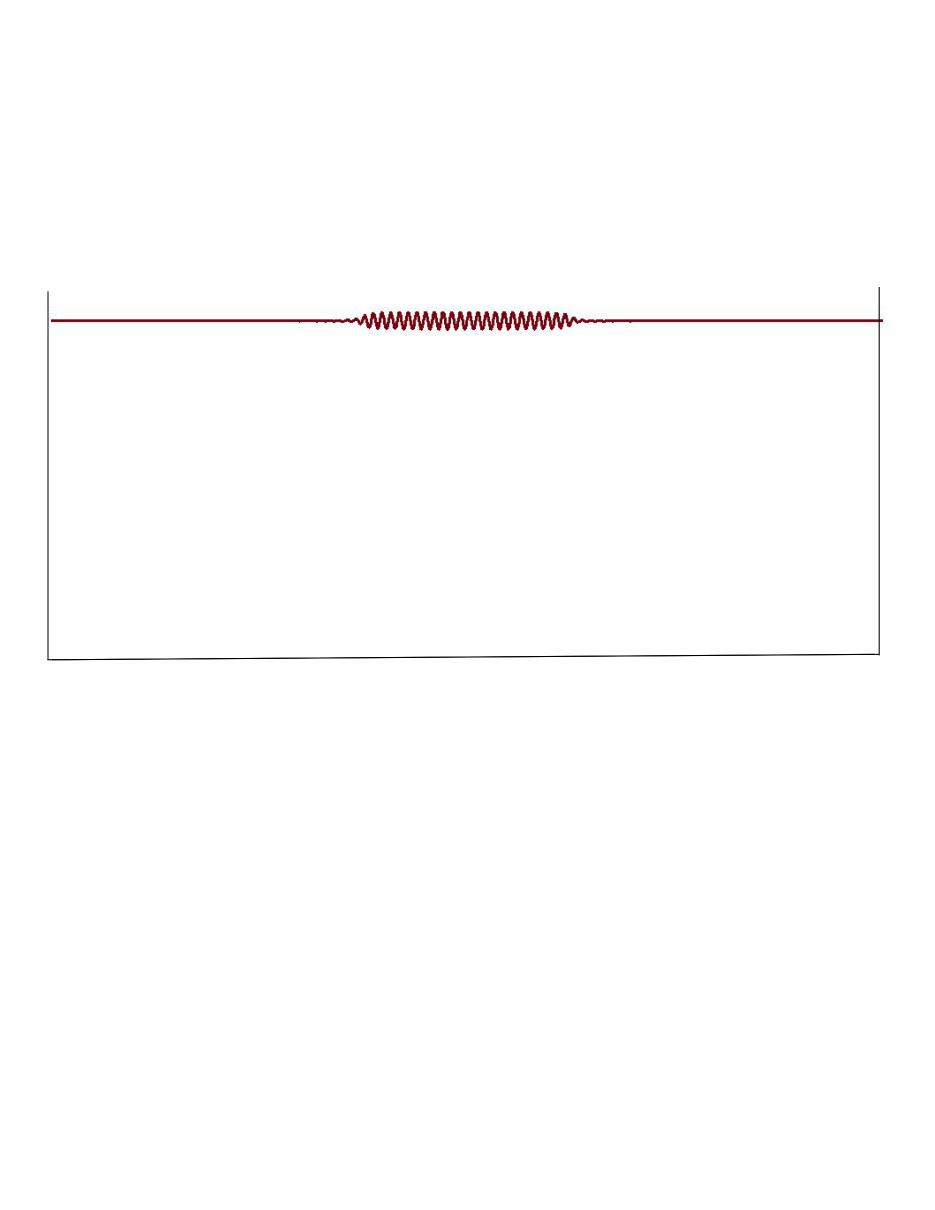} 
   \caption{An isolated propagating packet traversing a container.  }
   \label{localpacket}
\end{figure}

\begin{figure}[!h]
   \centering
   \includegraphics[width=4in,trim=0mm 0mm 0mm 0mm,clip]{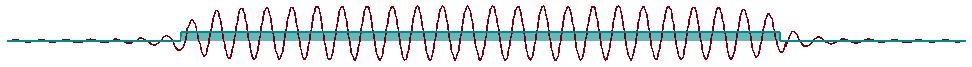} 
   \caption{Detail of the elevated wavepacket with wave height and elevation exaggerated elevation indicated by a superimposed shaded rectangle.  }
   \label{raisedpacket}
\end{figure}

We have calculated the mass flux (momentum density) above as $\frac{1}{2}\rho a^{2}\omega$.  For an incompressible fluid the presence of a mass flux in a packet requires the entire packet be elevated or the ends of the packet start to rise and sink respectively.  An infinite wave train obscures this fact.  We could consider a wavetrain moving around a looping tank with no ends and have no such situation since the total mass of the fluid must be conserved.  Thus it seems that to build a packet on our Airy basis of plane wave functions we must obtain a net elevation or end effects that our basis cannot easily provide.  The resolution to this apparent inconsistency is nonlinear by the way of the Poisson source term in the pressure constraint equation: $\nabla^{2}P=-\rho\partial_{i}v_{j}\partial_{j}v_{i}$.  Consider a rightwards propagating localized packet as in fig.\ \ref{localpacket}.  Anticipating that this wavetrain will generate a long range pressure field that washes out the small scale variation in the wave, let us consider a regional averaging of the source term on the right.  The time averaged local density on the right found by averaging over depth and time
\begin{align}
<\rho\partial_{i}v_{j}\partial_{j}v_{i}>=\frac{1}{2}\rho g a^{2}k /A
\end{align}
where $(A k^{-1})$ is the surface volume where most of the wave motion occurs.  The sign of the source term indicates that the pressure is locally increased due to the net negative ``pressure charge,'' $\rho_{P}$, in $\nabla^{2}P=\rho_{P}$ that is smeared out across the surface in the support of the wave.  The pressure field of the Airy wave is given by the linearized Bernoulli equation, $\rho\partial_{t}(a\frac{\omega}{k} \phi_{k})+P_{w}=0$ so that $P_{w}=\rho g a e^{kz}\sin(kx-\omega t)$.  The solution can be modified by elevating the water level in the support of the wave as in fig.\ \ref{raisedpacket} so that $e^{kz}\rightarrow e^{k(z-h)}$ where $h=a^{2}k$ so give the corrected solution for the new pressure.  If we now consider that the momentum of the packet must be given by $\rho h' v_{g}$ we see that $h'=a^{2}k=h$ so that the elevation of the packet from the nonlinear pressure correction gives the correct mass flux of the packet.  The correction from this elevated pressure will extend over the full walls of the container.  In the case of a standing wave that traverses the full container we have no net elevation by volume conservation of the liquid and the background pressure $P_{eq}$ must be adjusted to give the original net force on the bottom of the container which is independent of the presence of waves.\footnote{It is interesting to compare this analysis with the supposition that there is some irrotational flow carrying mass from front to back in a wavepacket as in McIntyre \cite{McIntyre} so that no net Stokes drift arises.  Data has refuted this model \cite{Smith} but it is comforting to have theoretical resolution for why that model does not occur.}  

\begin{figure}[!h]
   \centering
   \includegraphics[width=4in,trim=0mm 80mm 0mm 80mm,clip]{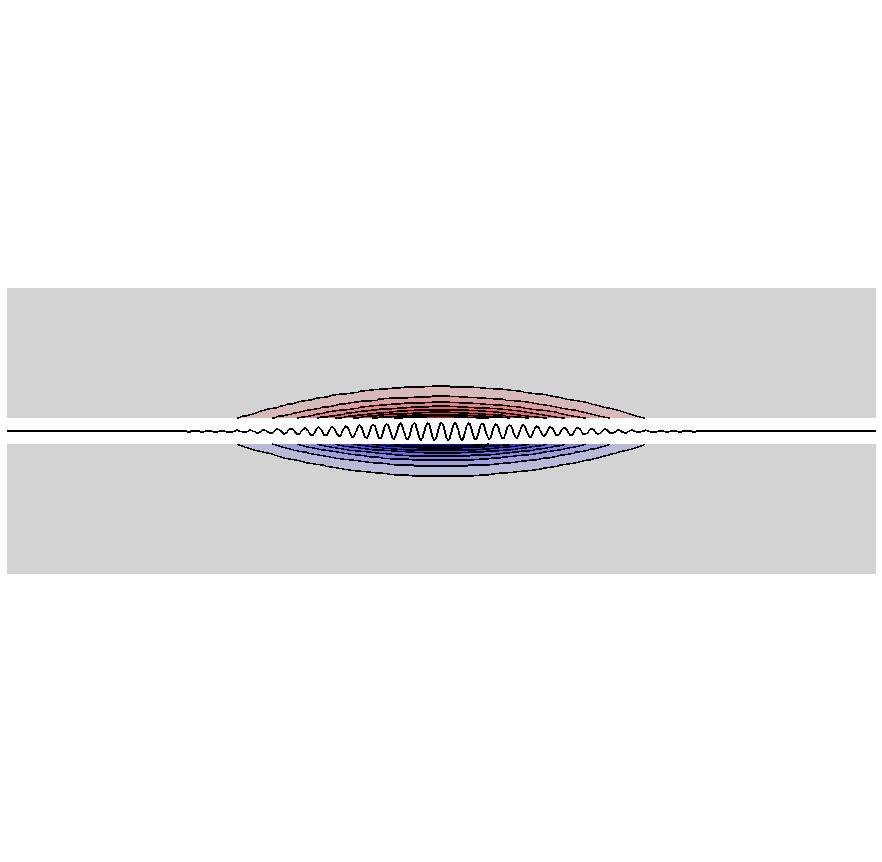} 
   \caption{A parallel plate construction with an image charge ``anti-advective'' flow to generate the surface forces from the waves.  The curves shown indicate isobars from the nonlinear contributions to the pressure.  }
   \label{plates}
\end{figure}

Another way to see this is to consider the case of initial data with an unelevated packet at time $t=0$.  To get the surface pressure equals zero we have a parallel plate problem with a negative ``pressure charge'' layer above the surface as an ``image charge'' (potentially due to an ``anti-advective'' flow with $Dv=\partial_{t}v-v\cdot\nabla v$) as shown in fig.\ \ref{plates}.  By the usual parallel plate analysis we see that the force density at the surface is $<f>=\sigma=\rho a^{2}\omega^{2}k$ so that the necessary equilibrium locally averaged elevation is $h\approx a^{2}k$.  

Results of Smith \cite{Smith} have shown that wave groups in an ambient sea can give negative net flow due to an Eulerian flow that exceeds the Stokes drift.  By mass conservation this means that a net surface depression is carried with the waves.  This model does not explain this but since the net mass flow is determined by such tiny elevation changes over the length of the wave groups, it is worth considering that some small contribution to the deep water flow on this larger scale is missing.  

\section{The Paradox of Wave Stress}
The forces on rigid bodies from surface waves is an important topic from an engineering and oceanographic perspective.  One of the classic problems is that of wave-set up whereby waves deposit extra water on beaches to raise the level of the mean surface.  This is often discussed in terms of ``radiative stress'' \cite{LH64} of the wavetrain.  The introduction of such a concept should raise some red flags to a hydrodynamicist.  Inviscid fluids do not support stress.  All the forces are mediated by pressure.  In the case of Reynolds stress on investigates a periodic wave motion and forms a time average.  The resulting motion for the averaged flow $v'$ involves the averaged pressure and a stress term that is independent of viscosity and models the nonlinear feature of the flow.  For a fluid with a free surface, we have seen that elevation changes in the surface can be important.  These are typically neglected in the motion of packets and it is presumed there is a flat/unelevated surface when averaged over a few wavelengths.  Ultimately, the forces imparted to walls are the result of a pressure field not stress.  Given the long history of confusion related to pseudomomentum \cite{McIntyre} and the usefulness in deriving real forces it seems prudent to consider this carefully.  

\begin{figure}
\centering
\begin{minipage}{2.2in}
  \centering
  \includegraphics[width=2.2in]{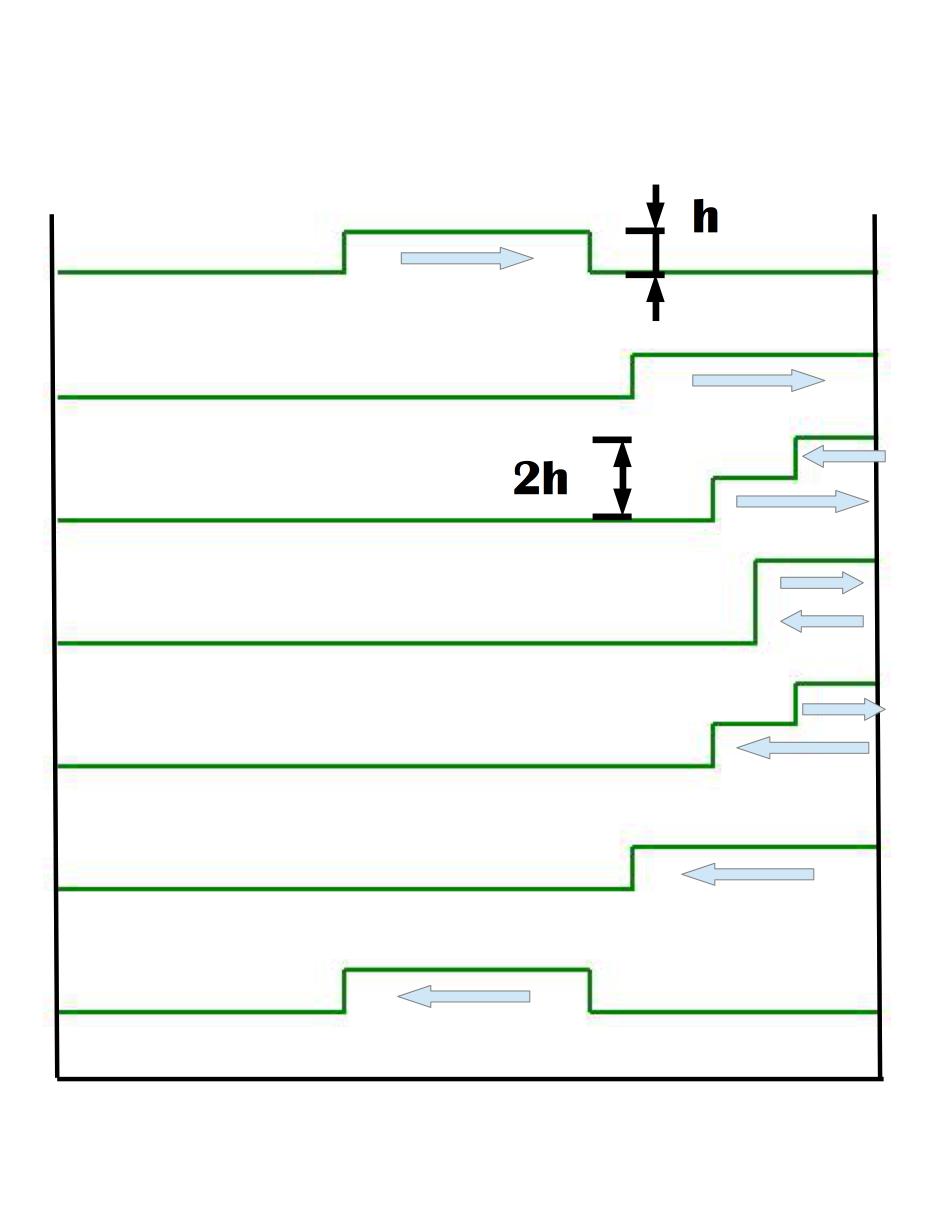}
  \captionof{figure}{Naive reflection of a packet incident on a wall with surface elevation effects labeled. }
  \label{sequence}
\end{minipage}
\hspace{1cm}
\begin{minipage}{2.2in}
  \centering
  \includegraphics[width=2.2in]{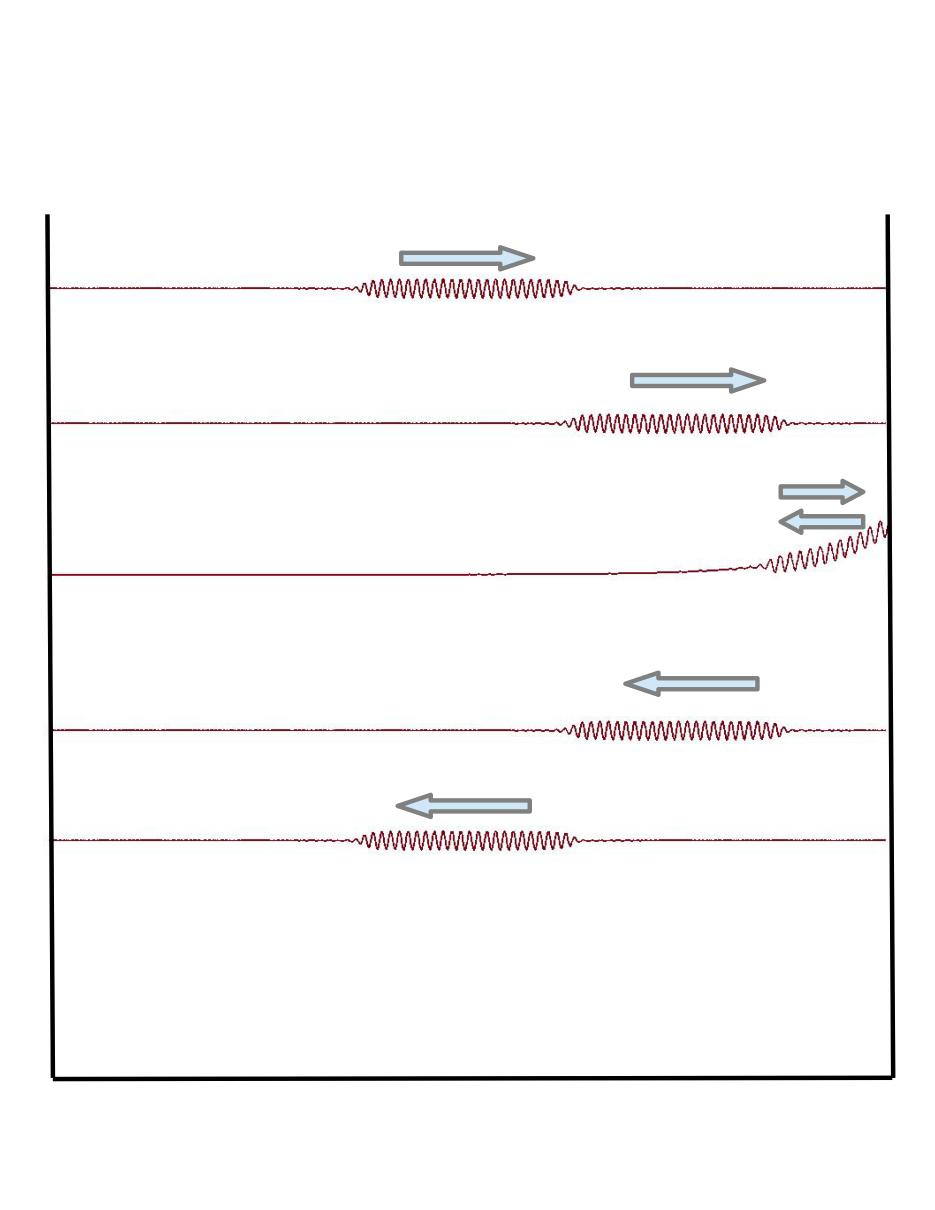}
  \captionof{figure}{A reflecting packet with transient elevation at the ends sufficient to transfer the momentum to the walls.}
  \label{pulses}
\end{minipage}
\end{figure}

Let us investigate two extreme cases.  First, that of of a localized packet as in fig.\ \ref{localpacket} and the case of a uniform standing wave in the same container.  In the case of a standing wave packet, it can be thought of as made of two oppositely moving propagating ones and the ends separate at $\sim\pm v_{g}$.  If the initial packet is made of net surface elevation $2h=a^{2}k$ then as the two waves gradually separate into two independent propagating waves of elevation $h$ similar to what is shown in the middle stages of fig.\ \ref{sequence}.  A propagating wave undergoes spreading from dispersion and spreads at the slower rate characterized by $k\partial^{2}\omega/\partial k^{2}$.  Consider packets long and fast enough so that the spreading induced by this is small over the timescale of the following thought experiments.  

In the case of a uniform standing wave, the pressure on the ends are generated by the static pressure and the contribution from the wave motion.  On a scale below the wave depth this is found from $\nabla^{2}P=\rho_{P}$.  Since this is a uniform time averaged distribution we have the same pressures on the ends of the container as on the sides and it averages to the static case.  Often it is said that the ``wave stress'' exists in the region of $\pm a$ of the surface.  If this is so, and there is a different in the time average longitudinal and transverse pressure on the walls, it gives a discontinuity at the corners of the container which is not reasonable since all the forces must ultimately come from a pressure.  

Now let us consider a packet advancing to the left as a propagating wave and being reflected.  We could naively consider the packet to be incident on the wall and reflected as in fig.\ \ref{sequence}.  The elevations have been exaggerated for illustration.  The height of the packet is $2h$ at its peak the force on the wall can be only $\frac{1}{2}\rho g (2h)^{2}$.  
If we compute the time averaged forces on the walls of a container holding a standing wave then the Stokes drift force $F_{s}=2 \rho h v_{g}=\rho g a^{2}$ gives the impulse per length of wall.  To maintain such a force the ends of the packet must be elevated at the wall by an amount $H=\sqrt{2}a\gg h$.  This result is illustrated in fig.\ \ref{pulses}.  The lateral forces on the container are given by a similar pressure field but one that is localized to the ends of the container.   
The resolution of the paradox is in that the long range pressure fields of the waves can create surface elevation changes that provide the necessary balancing forces that are not evident from any local linear analysis of the wavetrain or any Stokes expansion terms.

\section{The Effect of Shear Flow on Waves}
Armed with a greater confidence that we understand how waves generate momentum and pressures at the surface let us consider the influence of surface shears on waves from the standpoint of conservation laws.  
The stability of wave groups requires that the shape of the waveform be preserved.  One can have wave packets that do not spread where the group and phase velocities differ as long as the higher dispersion terms $\partial^{n}\omega/\partial k^{n}$ for $n>2$ all vanish.  Rogue waves and persistent wave groups \cite{Smith} seem to persist in their shape.  The very existence of the ``three sisters'' suggests that this is so.  We know of one situation where this automatically occurs: shallow water waves.  Let us consider how shallow water waves support solitons and other nondispersive waveforms of persistent shape.  We can do this from two points of view: energy flux and mechanical end forces.  

Consider a deep water wave.  The $\lambda$-scale averaged energy density $\mathcal{E}$ is evenly divided between kinetic $\mathcal{K}$ and potential energy density $\mathcal{U}$.  Since very little drift occurs compared to local circulation it is an excellent approximation to say that the PE advances at the phase velocity $v_{ph}$ and the KE stays fixed \cite{Chafin-EM}.  The front end of the wave packet must convert the elevation changes to kinetic motion to continue the wave.  This picture explains why $\mathcal{U}v_{ph}=\mathcal{E}v_{g}$.  In the shallow water case there is no dispersion.  Almost all the energy is potential so $\mathcal{U}\approx\mathcal{E}$ and $v_{ph}=v_{g}$.  This picture works very generally for all wave phenomena in media when one considers a decomposition of the part that is resting and the part that is traveling.\footnote{For a dielectric analog see Chafin \cite{Chafin-EM}}  

\begin{figure}[!h]
   \centering
   \includegraphics[width=4in,trim=0mm 30mm 0mm 0mm,clip]{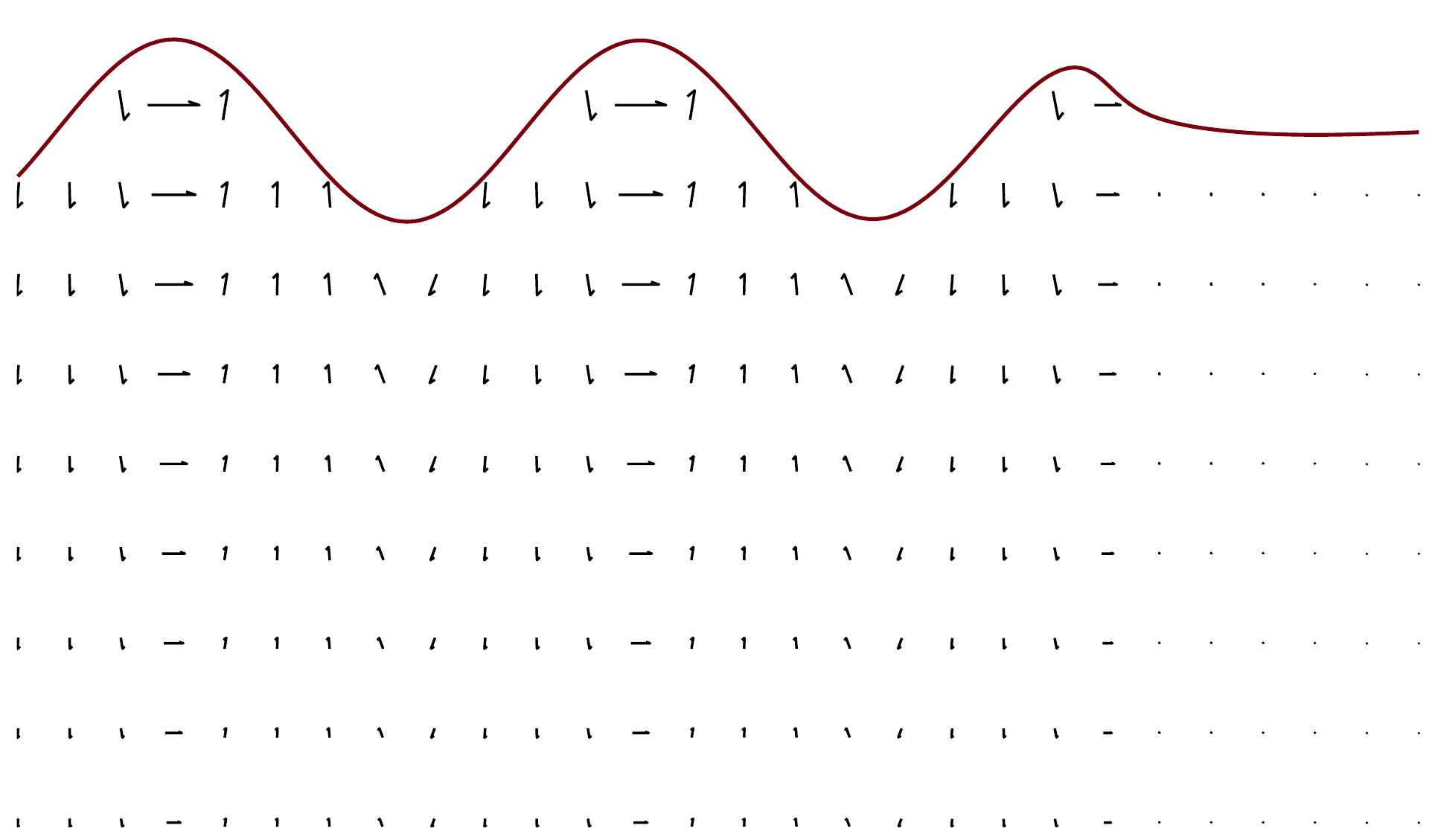} 
   \caption{Advancing edge of wavepacket.}
   \label{front}
\end{figure}

Let us now consider this from a mechanical force perspective.  The ends of a packet spread because the ends of the packet produce fields at a distance longer than $\lambda$ since there are no adjacent wavecrests to balance it as in the interior of the packet as in fig. \ref{front}.  In a shallow water wave the pressure field extends down to $\sim\lambda$ in depth but most of this is into the solid material that forms the bottom of the lake or seabed.  The only part that can drive up waves away from the ends are more local to the crests themselves.  This is why there is no dispersion in shallow water waves and persistent elevations like solitons can exist.  
Above we made the distinction between the forces exerted by Stokes drift and the spreading of a propagating packet itself.  We are interested in the latter.  In particular, we want to know how currents can balance these end forces to prevent dispersive spreading.  

First let us consider the case of a deep discontinuous shear flow.  Assume the surface layer of the ocean down to a depth $a<<d<<\lambda$ move at a constant velocity over the lower surface.  If this flow is faster than the the waves move it effectively a solid surface on which waves behave as shallow water waves.  Specifically, if the rate of the flow $v_{f}$ gives a period $\Omega=v_{f}\lambda\ll \omega$ the deep layers of water don't have time to respond and contribute to irrotational motion.  Surface flows tend to be rather small compared to the speed of large waves so this example is not our main interest but it may play a role for small waves in rapidly increasing winds.

\begin{figure}[!h]
   \centering
   \includegraphics[width=4in,trim=0mm 120mm 0mm 120mm,clip]{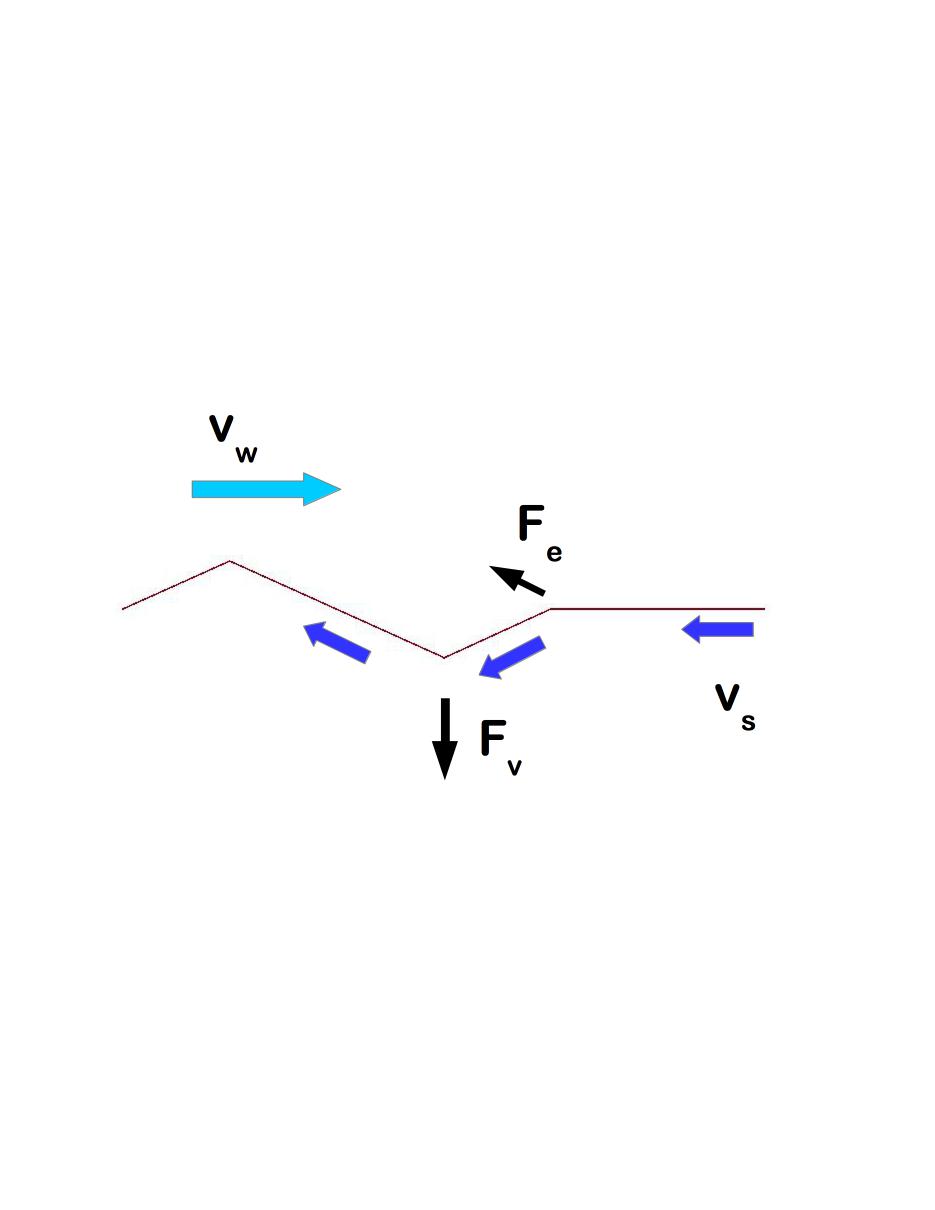} 
   \caption{A right moving wavepacket moving at velocity $v_{w}$ with incident thin surface shear moving at $v_{s}$.  The impulses $F_{e}$ at the end and $F_{v}$ act at then packet ends and vertically at the peaks and troughs respectively. }
   \label{forces}
\end{figure}

On the ocean, waves often have wavelengths in the 10's of meters.  The surface shear generated in storms can be only a few meters deep so we have a situation where the surface shear can act as a shallow modification of the waves.  For our model consider a narrow flow of depth $\delta$ and depth integrated surface density $\sigma$ of water that travels over our waves.  Vorticity conservation requires that the surface shear be the same at all points along the surface regardless of the height of the wave.  This necessitates that the horizontal component of its velocity be reduced on the inclined portions.  The advancing edge of a packet is shown in fig.\ \ref{front} with velocity field superimposed.  For simplicity  we have use a triangular wave to illustrate the flow and forces in fig.\ \ref{forces}.

The amplitude of this wave is $a$ and the wavelength $\lambda$ so the angle of rise at the end of the packet is $\theta=\arctan(4a/\lambda)$.  We now wish to calculate the impulse due to the incident flow but we have to remember that the flow represented by $\sigma v_{s}$ corresponds to fluid that would be considered as part of the wave solution so that as $v_{s}\rightarrow0$ the impulse vanishes.  The change in the x-component of the flow is found geometrically by mass and vorticity conservation to be $\Delta v_{x}=v_{s}(\cos(\theta)-1)$.  The impulse on the end is given by the rate its momentum arrives at the packet edge $u=v_{w}-v_{s}$ (where we have assumed the positive direction of both flows is rightwards).  The impulse (force per length) on the packet is then $F_{e}^{x}=f=\sigma u \Delta v_{x}$.  This is an inwards force and it is returned when the flow exits the packet leading to a stress on the packet the packet of $S_{xx}=f$.  
We know that the outwards force on the packet edge from elevation is $\frac{1}{2}\rho g h^{2}$.  Since the elevation is a function of the pressure induced by local wave motion, this tiny force is all that needs to be matched to prevent packet spreading.  

There is a secondary effect of this flow.  The vertical forces $F_{v}$ decrease the restoring pressures of the wave elevation $\eta(x)$ and allow waves of the same wavelength to grow larger.  This force is analogous the the radially expanding force on a coiled flexible hose when water flows through it.  For our shallow wave approximation we can estimate this force to be $F_{v}\approx2\sigma u v_{s}\sin(\theta)/\lambda$.  Compared to the pressure $P_{w}(z=0)\approx \rho g a$ from the waves for these to be comparable we need surface flows much faster than the motion of water in the waves themselves $\sim a\omega$.  As the waves become steeper this force becomes more appreciable and provides an avenue to compress the energy of waves into shorter packets of larger waves.  

Neglecting this vertical force we can use the usual dispersion relation for Airy waves and compute the shear flow necessary to confine the packet.  Using our triangle wave model we can find the shear velocity necessary to keep a packet of a given wavelength and amplitude bound.  Using that $h=a^{2}k$ we have the equilibration condition 
\begin{align}
\frac{1}{2}\rho g a^{2} (ak)^{2}=\sigma u \Delta v_{x}
\end{align}
The surface density is $\delta\le a$ thick so we have $\sigma=\rho\delta$.  Using the small angle correction $\cos(\theta)\approx1+\frac{2}{\pi^{2}}(ak)^{2}$ it follows
\begin{align}
\frac{\pi^{2}}{4}g a^{2}\approx \delta u v_{s}
\end{align}
The velocity of long waves is much faster than surface currents so $v_{s}\ll v_{g}=\frac{1}{2}\sqrt{\frac{g}{k}}$.  Using the dispersion relation $\omega=\sqrt{gk}$ we have
\begin{align}
v_{s}\approx\frac{\pi^{2}}{2}a\omega \frac{a}{\delta}
\end{align}
where $a\omega$ is the maximum velocity of irrotational flow at the surface.  For a surface shear of 1m/s of 10s waves we have $a (a/\delta)\doteq0.32$.  Thus a shear depth of three meters is enough to stabilize a packet with amplitude 1m.  Steeper waves will lead to stronger forces inwards forces where the small angle approximation fails.  Since the wave motion falls off with length $\lambda$ the approximation here is good for $\delta\ll \lambda$ even when $\delta>a$.  In general, such stabilization is possible when there is a surface shear depth comparable to the amplitude of the wave and the fluid velocity of the wave $a\omega$ is less than the surface shear velocity $v_{s}$.  

This analysis shows that this stabilization can occur for wave in the ``linear regime,'' by which we mean that Airy wave theory works well.  It is often assumed that rogue wave phenomena is a nonlinear effect and, to the extent that it applies to waves with $ak$ that are not small this is obviously true.  The question one should ask is whether rogue waves are to be defined as waves notably larger than the ambient ones or as very steep ones.  The latter case is nonlinear but not necessarily in an interesting way.  The former definition implies that shear and waves can give ``rogue'' waves by linear analysis.  It was only by considering the small nonlinear pressure terms we could derive this.  As usual, nonlinearity is always tied up somewhere in any interesting wave phenomenon.  When one considers steeper waves there is a cycloidal distortion that steepens the peaks and broadens the troughs.  This reduces the Stokes drift proportionately compared to a low amplitude wave.  The drift gives a measure of the elevation necessary for the packet so we expect the shear flow intensity necessary for localization to decrease rather than grow for larger waves.  

\section{Conclusions}
Conservation laws are often a powerful tool for solving problems when exact dynamical solutions are elusive and perturbative methods are uncertain to capture the desired effects.  For continuum systems the subtleties in such an approach are surprising and the use of wavepacket analysis is essential.  We have seen that the waves create a time averaged net positive ``pressure charge'' source that elevates the packets just enough to explain the Stokes drift without cumulative and lossy end-of-packet elevation changes.  The induced stress of this for propagating waves is shown to be surprisingly easily balanced with surface flows but this requires a more subtle analysis of the kinds of stresses that appear in the ``radiation stress'' of waves used in usual calculations of wave set-up on beaches.  

One might wonder if the ambient waves themselves could generate such a mass flux by the Stokes drift to stabilize the packets.  Small waves can be destructively acted on by larger ones \cite{Ha} so the question requires a deeper analysis.  Since real world waves typically carry negative Eulerian mass flux to cancel the Stokes drift \cite{Smith} the relative depth of it will be important.  We have only investigated stability of the packets not questions of generation of them and how rogue waves build to such large amplitudes.  Nonlinear effects almost certainly play a role in this.  

There are many aspects of rogue waves that require further elucidation.  Since the elevation changes are over the scale of whole packets, it is not possible that perturbative effects can capture this mechanism.  It is astoundingly easy to use perturbative methods to generate nonlinearities that lead to soliton behavior but it is often unclear if this is preserved by higher order changes.  Furthermore, as we saw in the case of angular momentum, infinite wave analysis cannot always be a faithful guide to give conserved quantities and net forces on packets.  

Even if one accepts that shear flow is mechanically stabilizing the waves, it is unclear how universal it is in stabilizing and generating unusually large waves.  Ambient waves can carry mass as well and produce their own forces as they travel over, and possibly get destroyed by, larger waves. The ``three sisters'' phenomenon for rogue waves suggests that the shear stress should be uniform across the waves so there should be some constraints on the relative lengths and amplitudes among this set of three waves.  We have not included the vertical component of momentum from the shear and the back interaction of the wave on it.  It would be interesting if this led to some means of energy extraction from the flow to build wave amplitude by introducing vorticity so that the momentum of the flow remains the same while its kinetic energy decreases.  

\section*{Acknowledgements}
Thanks to Renee Foster and Bo Hemphill for proofing this and earlier versions.  Special thanks to Thomas Sch\"{a}fer for thoughtful discussions.

\end{document}